\documentclass{llncs}
\usepackage{epsfig}
\usepackage{makeidx}  
\begin{document}
\frontmatter          
\pagestyle{headings}  
\pagenumbering{arabic}
\title{Interning Ground Terms in XSB}
\titlerunning{Interning Ground Terms in XSB}  
%
\author{DAVID S. WARREN
        \thanks{Research supported in part by National Science Foundation, Award 0964196}
}
\authorrunning{D. S. Warren}   
%
\tocauthor{David S. Warren}
\institute{Stony Brook University, Stony Brook, NY 11794-4400, USA,\\
\email{warren@stonybrook.edu}
}

\maketitle              

\begin{abstract}

This paper presents an implementation of interning of ground terms in
the XSB Tabled Prolog system.  This is related to the idea of
``hash-consing''.  I describe the concept of interning atoms and
discuss the issues around interning ground structured terms,
motivating why tabling Prolog systems may change the cost-benefit
tradeoffs from those of traditional Prolog systems.  I describe the
details of the implementation of interning ground terms in the XSB
Tabled Prolog System and show some of its performance properties.
This implementation achieves the effects of that of Zhou and Have
\cite{zhou-hash-consing} but is tuned for XSB's representations and is
arguably simpler.
\end{abstract}
%
\section{Introduction}

Prolog implementations (and all implementations of functional
languages that I know of) intern atomic constants.  An atomic constant
(called an ``atom'' in Prolog) is determined by the character string
that constitutes its name.  Rather than representing each occurrence
of an atom by its character string name, the character strings are
kept uniquely in a global table and the atom is represented by a
pointer to its string in that table.  This (usually) saves space in
that multiple occurrences of the same atom are represented by multiple
occurrences of a pointer rather than multiple occurrences of its
string.  But more importantly, comparison of atoms is simplified; two
atoms are the same if and only if their pointers are the same.  The
important direction here is that, since each string appears only once
in the global table, two atoms differ if their pointers differ.  This
makes atom comparison simpler and more efficient.

The atom table is indexed, usually by a hash index, so finding whether
a new atom already exists in the table (and adding it if it does not)
is a relatively efficient operation.  This operation is known as
``interning'', indicating that an atom representation is converted
from a string representation to an internal representation, i.e., the
pointer representation.

The question arises as to whether this kind of representation might be
lifted to more complex terms, i.e., applied not only to atoms but to
structured terms.  This idea has been explored in the Lisp language
community \cite{goto74} and goes by the name of ``hash-consing''
(originally proposed by \cite{ershov}).  The name comes from the use
of a ``hash'' table to store the structures, and in Lisp the way that
complex structures are constructed is by means of an operation called
``cons''.  Zhou and Have
\cite{zhou-hash-consing} present an implementation of this concept
in B-Prolog.  I compare my approach to theirs in more detail later
in the paper.

There are several reasons why interning of structured terms is more
complex than interning atoms, and its potential advantages over a
traditional direct implementation of structured terms less clear.  The
interning operation itself, while optimized by sophisticated indexing
strategies, still takes time.  Interning of atoms is required when a
new atom is created.  Atoms are created (mostly) at read-in time, and
sometimes in particular builtins such as atom\_codes/2.  Execution of
pure code (e.g., without builtins) does not cause the creation of new
atoms so the overhead of interning atoms is relatively small and
localized in most Prolog programs.  However, structured terms are
created continuously during execution of pure code.  I assume a
``copy'' based implementation of terms, which is the implementation in
WAM-based Prolog systems.  For example, running the traditional
definition of append/3 to concatenate two lists requires the creation
of as many structured subterms as there are elements in the first
list.  So the interning cost for complex terms can be quite high.
Significant memory may be saved by interning structured terms, but the
space-time tradeoffs are not clear.  Many terms may be created and
used once but then not used again.  Locality of reference is also
changed, so caching behavior may be affected, perhaps for the better,
perhaps for the worse.

There is another complication in the case of Prolog that does not
arise in functional programming systems: Prolog terms may have
variables embedded in them.  Interning a term containing a variable is
problematic.  For example terms, such as f(a,X) and f(a,Y), cannot be
interned to the same hash table entry, since X may become bound to b
and Y to c, in which case the terms are distinct.  However, if they
are interned to distinct table entries, then if X and Y both become
bound to a, the terms are then the same but in this situation there
are two distinct copies of the same term in the hash table, which
undermines a major reason for interning.  These (and other)
difficulties strongly mitigate against trying to intern terms in
Prolog that contain variables, i.e., terms that are not ground.  (See
the recent work of Nguyen and Demoen \cite{nguyen-demoen-rep-sharing}
for an interesting, and deeper, discussion of this issue.)

One might explore interning every term when it is created.  In the
WAM, terms can be created in a variety of ways.  In pure code (i.e.,
not in builtins) terms are built in a top-down way by a sequence of
instructions starting with a get-structure instruction and followed by
a sequence of unify-something type instructions, one for each field of
the term.  These instructions could be changed to support checking
whether all subfields contain constants or interned subterms, and to
intern the constructed term if so.  But this would require major
surgery to these instructions.  A better solution would probably be to
modify the WAM instructions and compilation strategy to build terms
bottom-up.  But again, it is not clear, even
with such optimizations, that the overhead of hashing every time a
ground term is constructed would be out-weighed by other improvements.

For all these reasons, I believe, general interning of ground complex
terms is not a general implementation strategy considered in Prolog
systems.

The advent of Prolog systems that support tabling, however, may have
changed the cost-benefit analysis of sometimes interning ground terms.
When a tabled predicate is called with new arguments, the arguments of
the call are copied into the table; and when new answers are returned
to a tabled predicate, they are also copied into the table.  Answers
are also copied out of tables when they are used to satisfy subsequent
calls.  This copying of calls and answers to and from tables may lead
to significant time and table space usage.  For example, when using a
DCG (Definite Clause Grammar) in the standard way for parsing, the
input string (represented as a list) is passed into each nonterminal
predicate and the list remaining after the nonterminal has recognized
a prefix is passed out.  So, for example, when tabling a nonterminal
predicate that removes just the first atom in a long list, the entire
list is copied into the table once for the call, and the entire list
minus the first element is (again) copied into the table as the
answer.  (The fact that tries are used to represent calls and answers
in tables may in special cases reduce the copying, but in general, it
is needed.)  So when using DCG's to parse lists of terminal symbols,
there is much copying of lists into and out of tables.  Tabling a DCG
can in principle give the performance of Earley recognition, but this
extensive copying of the input list adds an extra ``unnecessary''
linear factor to the complexity, in both space and in time.

Note also that when a term is copied into the table, it must, of
course, be traversed.  So it adds no extra complexity to check for its
goundedness and intern it if it is ground.  Another situation in which
this happens is in assert.  Since an asserted term is fully traversed
to convert it into internal ``code form'' (in the XSB implementation
of assert), one can intern ground subterms during that process without
increasing complexity.  Another opportunity would be in findall/3.


\section{Implementation of Interned Ground Terms}

\subsection{Representation of Interned Ground Terms}

\begin{figure}[ht]
\begin{center}
\includegraphics[trim=0cm 8cm 0cm 0cm, clip=true,
totalheight=0.4\textheight]{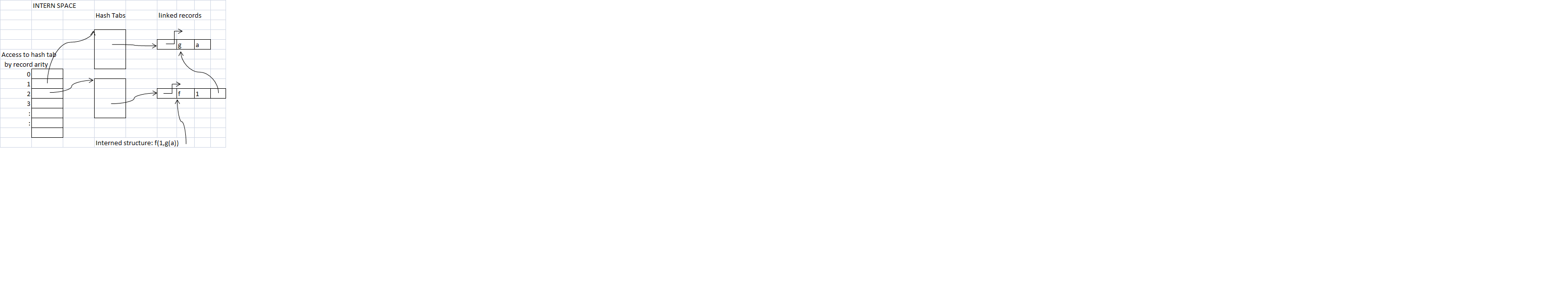}
\end{center}
    \caption{Storage of Interned Ground Terms\label{fig:internspace}}
\end{figure}

I describe the representation of interned ground terms in XSB.  In the
WAM structured records are represented as a sequence of words, the
first is a pointer to a global record for the function (aka structure)
symbol.  For a structure symbol of arity $n$, that initial word is
followed immediately by $n$ tagged words representing the subfields of
that structure.  List (or cons) records are just pairs, with the
structure symbol optimized away in favor of a tag.  A picture of a
portion of the state for interned terms, containing the ground term
f(1,g(a)), is shown in Figure \ref{fig:internspace}.

Interned structure records of arity $n$ are stored in blocks of
records, each of $n+2$ words, ``linked records'' in the figure.  The
records are accessed (for interning) by using the record arity to
index into an array to access a hash table for records of that arity.
The hash value is computed using the $n+1$ fields of the record.  The
subfields of an interned record can contain only atoms, numbers, or
tagged pointers to other interned records, and so the hash value
computed from these values will be canonical.  The hash value is used
to index into the hash table to access a hash bucket chain that can be
run to find the desired record.  The bucket link field immediately
precedes its record.  List (or cons) records have their own special
hash table.

The representation of interned terms is exactly the same as in the
heap; the only difference is that the records are stored in globally
allocated blocks, not in the heap.  For example, in the Figure, the
pointer from the bottom to the f/2 record could well be from the heap,
and for any code traversing this representation, the data structure
looks exactly the same as it would were in on the heap. This means
that all existing code in XSB for accessing and processing structured
terms continues to work with interned ground terms.

Whether a structured term pointer is pointing to an interned term or
not is determined by examining the pointer itself; if it points into
the heap, it is not interned; otherwise it is interned.  One can think
of this as adding another ``tag'' to a pointer to a structured term, but
the ``tag'' is implemented using a pointer range, rather than an
explicit bit in the pointer.  The general unification algorithm is
modified to check, when unifying two structured terms, if the terms
are both interned, in which case it fails if the addresses are not
equal.  (Note that the algorithm already succeeds immediately if two
pointers to structure records are equal.)

Other builtin functions can be modified to take advantage of knowing a
subterm is interned.  For example, the builtin ground/2, which checks
for groundedness, need not descend into an interned subterm; the
builtin copy\_term/2 does not need to descend into an interned ground
term but can simply copy the reference.

\subsection{Interning Ground Terms}
A new function (accessible through a builtin) takes a Prolog data
object (usually a structured term) and creates a copy of that term in
which all ground subterms are interned.  The term is traversed
bottom-up, using an explicit stack, and the new copy is created on the
heap.  (Of course, if the term is ground, the new heap copy will be a
single word pointing the interned term.)  Clearly subterms that are
already interned need not be traversed; the reference to the existing
interned representation is simply copied.  Note this operation is
different from the standard copy\_term/2, since the new term contains
the same variables as the old term, whereas in copy\_term/2 the new
term contains new variables.

The user can call the builtin intern\_term/2 at any time to make a
logically identical copy of any term.  Since interned ground terms are
represented exactly as regular heap terms, except that they reside in
a different place in memory, nothing in the XSB system needs to be
changed to support the terms created by the new builtin
intern\_term/2\footnote{In fact, in XSB the builtin findall/3 copies
terms out of the heap and uses the fact that term pointers do or do
not point into the heap to determine sharing.  Therefore the
distinction between findall/3 terms not in the heap and interned terms
not in the heap had to be handled carefully within this operation.}.
However to take full advantage of the new term storage mechanism,
other system changes can be made.  I describe changes made to
asserting of dynamic code and to the handling of complex terms in
tables.

\subsection{Interning before Asserting}

Terms are fully traversed in XSB when a clause is asserted to the
generate WAM code that will be executed when the clause is called.
Also, when that code is called, it may construct a copy of a term on
the heap.  Thus interned terms can be used to good effect when
asserting clauses.  New WAM instructions are added for
get-intern-structure (and unify-intern-structure) whose (non-register)
argument is a pointer to an interned term.  If a dynamic predicate is
declared as intern, then clauses are automatically interned before
they are asserted.  This can save space if the same ground term occurs
multiple time in asserted clauses.  Again, this doesn't increase the
complexity of assert, since the clause has to be fully traversed in
any case.

Get-intern-structure unifies a term with an interned term.  Unlike
get-structure, it will never construct any subterm on the heap, since
all subterms of the interned term are interned and unifying an
interned term with a variable simply sets the variable to point to the
interned term.  So this can save space on the heap and the time it
would otherwise take to construct the term on the heap.

Note that indexing is not an issue with asserted clauses in XSB, if
they are not trie-indexed.  Standard hash indexing still hashes on the
same portion of an argument term, whether it is interned or not.
Based on the resulting hash value, it chooses the set of clauses that
might unify, and then executes the chosen clauses that actually do the
unification.

\subsection{Interning before Tabling}\label{int-bef-tab}
The computational advantage of interned terms is that the system never
needs to make a copy of one; it can simply use its reference.  As
described above, terms are copied into and out of tables to represent
calls and answers.  With interned subterms much of this copying can be
avoided.

\begin{figure}[ht]
\begin{center}
\includegraphics[trim=0cm 0cm 0cm 0cm, clip=true,height=0.2in,width=4.0in,
totalheight=0.4\textheight]{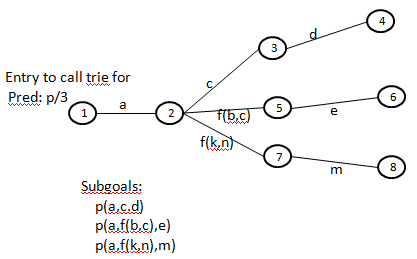}
\end{center}
    \caption{Trie Containing Interned Ground Terms\label{fig:interntrie}}
\end{figure}

In XSB a variant table can be declared as intern, in which case all
calls will be interned before being looked up, and possibly entered,
in the table.  Similarly, all returned answers will be interned before
being (checked and perhaps) added to the table. In XSB terms in a call
(and return) table are represented in tries, using a linearization
based on a pre-order traversal of the terms.  Figure
\ref{fig:interntrie} shows schematically a trie that contains 
interned ground terms.

The Figure shows a trie containing three calls to p/3: p(a,c,d),
p(a,f(b,c),e), and p(a,f(k,n),m), assuming that p/3 is declared as
intern.  The new feature here is that, for example, f(b,c) is a ground
complex term and so is interned before being entered into the trie.
So the entire interned subterm is treated as a unit and represented as
being on one link, such as between node 2 and node 5.  When an
interned subterm is encountered when adding (or looking) up a
component in such a trie, it is treated as an atomic constant, with
the reference treated as the unique identifier.  This figure shows a
(possibly complex) symbol on each link, but of course, in the
implementation that is a pointer to some canonical representation for
that symbol.  For a constant it is a pointer to the interned string of
its name; for an interned structured term, it is a pointer to the
canonical representation for that term in the interned term data
structure.

Notice that interning the arguments when making a tabled call does not
increase the complexity of processing the table, since the terms, were
they not interned, would have to be completely traversed in any case.
At worst, the constant factor may increase due to the multiple
traversals.

It is worth noting how interned terms interact with the indexed lookup
of calls (and answers) in the table.  Each node in the trie can be
indexed, so, for example, a hash index is built (as necessary) on the
outgoing links from a node to quickly move to the target node on the
right outgoing link.  For example, node 2 in the Figure would have a
hash table to quickly access nodes 3, 5, or 7.  Thus tries ordinarily
provide full indexing on every constant and function symbol in a term
being looked up.  However, as we have seen, interned terms are treated
as (unstructured) constants in the trie, and are indexed as constants.
This means that there is no indexing on the main function symbol (or
indeed any component) within an interned term.  So, for example, if a
call is made to p/3 of the form p(a,f(k,X),m), when trie traversal
reaches node 2, it cannot use the hash table to index at this point to
find quickly the one term (on the link to node 7) that matches.  
Since the input term is f(k,X) is not ground, it is not interned, and
the index, which is based on pointers to ground terms cannot be used.
Note that were p/3 not tabled as intern, the trie would have more
links, and the symbols f/2 and k could be used to index the
traversal.  But given that p/3 is tabled as intern, the only choice
would be to look at every outgoing link from 2, and see whether the
possibly complex symbol unifies with the lookup term.

So this loss of indexing may potentially have serious performance
consequences.  However, if only variants of the source term are to be
retrieved, and all ground subterms in both the source lookup term and
the trie are known to be interned, which is the case for variant table
processing, then this problem is avoided.  Note also that tables that
are not declared as intern will process interned terms just as they do
regular terms, traversing them and processing each atomic component.

The implementation in XSB currently avoids this potential problem by
disallowing the entry of interned terms into tries for which retrieval
by unification would be required.  This may be revisited, since there
do seem to be situations in which the benefits of interning could be
gained and the pitfalls of the loss of indexing avoided.

\section{Performance}

All tests were done on a laptop, running Windows 7 Professional,
64-bit OS, on a Intel(R) Core(TM i5) CPU, 2.67 GHz with 8 GB of
memory.  XSB was compiled using MSVC in 64-bit mode.

Figure \ref{intern-time-table} shows how long it takes to intern a
list of integers.
\begin{figure}
\begin{center}
\begin{tabular}{|r|c|}
\hline
List Length &    CPU Time (secs) \\ \hline
 100000     &     0.0160 \\
 200000     &     0.0470 \\
 300000     &     0.0630 \\
 400000     &     0.0940 \\
 500000     &     0.1090 \\
 600000     &     0.1400 \\
 700000     &     0.1720 \\
 800000     &     0.2030 \\
 900000     &     0.2180 \\
1000000     &     0.2340 \\ \hline
\end{tabular}
\caption{Time to intern a ground list}\label{intern-time-table}
\end{center}
\end{figure}
Each run starts with an empty intern table, so every new subterm must
be added.

A simple (but not very realistic) example in which interning can
provide great performance improvements (see \cite{zhou-hash-consing})
is to table a predicate that tests that a term is a proper list:
\begin{verbatim}
:- table islist/1 as intern.
islist([]).
islist([_|L]) :- islist(L).
\end{verbatim}
and call it with a list of distinct integers of various lengths.
Figure \ref{islist-table} shows the results.
\begin{figure}
\begin{center}
\begin{tabular}{|c|c|c|c|c|}
\hline
List Len &  nonintern &      nonintern  &     intern &      intern \\
         &   Cpu Time &     Table Space &    Cpu Time &    Table Space \\
	 &    (secs)  &        (bytes)  &      (secs) &      (bytes)\\ \hline
  100  &      0.0000  &           427,304  &   0.0000 &          27,264 \\
  800  &      0.0160  &        25,813,640  &   0.0000 &         213,600 \\
 2700  &      0.2490  &       292,321,384  &   0.0160 &         721,344 \\
 6400  &      1.3570  &     1,640,106,376  &   0.0000 &       1,722,720 \\
12500  &      5.8960  &     6,253,333,160  &   0.0150 &       3,333,120 \\ \hline
\end{tabular}
\caption{Interning of islist/2: Space and Time Comparisons}\label{islist-table}
\end{center}
\end{figure}
Without interning, each recursive call causes the sublist to be copied
into the table.  So every suffix of the initial input list is copied
to the table, and the space required is quadratic in the length of the
input list.  With interning, only a pointer to an interned list is
copied into the table, so the space required is linear in the length
of the input list.

DCGs normally process lists and can benefit significantly from
interned structures.  Consider the following DCG for a grammar that
recognizes even-length palindromes:
\begin{verbatim}
:- table epal/2.
epal --> [].
epal --> [X],epal,[X].
\end{verbatim}
Figure \ref{palindrome-table} shows the results of recognizing a list
of randomly chosen numbers between 1 and 10,000,000, appended to its
reverse, to make an even-length palindrome.
\begin{figure}
\begin{center}
\begin{tabular}{|c|c|c|c|c|} \hline
List Len &  nonintern &      nonintern  &     intern &      intern \\
         &   Cpu Time &     Table Space &    Cpu Time &    Table Space \\
	 &    (secs)  &        (bytes)  &      (secs) &      (bytes)\\ \hline
200      &  0.0000    &     3,642,296   &     0.0000  &      65,664 \\
1600      &   0.2340   &     230,735,064   &     0.0000  &      526,944 \\
5400     &   2.6060   &   2,625,534,840   &    0.0160   &     1,766,336  \\
12800     &   xx   &   xx   &  0.0310     &    4,213,088   \\
25000     &   xx   &   xx   &   0.0780     &  8,231,424   \\
43200    &   xx   &   xx   &  0.1090     &   14,259,296   \\
68600    &   xx   &   xx   &  0.1720     &   22,751,040    \\
102400    &   xx   &   xx   &  0.2960     &   34,226,528    \\
145800    &   xx   &   xx   &  0.4680     &   48,288,128    \\
200000    &   xx   &   xx   &  0.6560     &  65,848,928    \\ \hline
\end{tabular}
\caption{Palindrome (epal) DCG: Space and Time Comparisons}\label{palindrome-table}
\end{center}
\end{figure}
The xx's indicate instances that do not run due to memory limitations.
Note that with interning, palindrome recognition is linear in
time and space.

An example of a grammar for which tabling is required is the following
left-recursive grammar, which recognizes all strings consisting of just
the integers 1, 2, and 3.
\begin{verbatim}
:- table lr/2.
lr --> [].
lr --> lr,[1].
lr --> lr,[2].
lr --> lr,[3].
\end{verbatim}
Figure \ref{lrrec-table} shows the results of using this grammar to
recognize strings, when using and not using interning.  The strings
are lists of integers 1, 2, and 3 chosen randomly.
\begin{figure}
\begin{center}
\begin{tabular}{|c|c|c|c|c|} \hline
List Len &  nonintern &      nonintern  &     intern &      intern \\
         &   Cpu Time &     Table Space &    Cpu Time &    Table Space \\
	 &    (secs)  &        (bytes)  &      (secs) &      (bytes)\\ \hline
100      &  0.0000    &        388,248   &     0.0000  &     5,168  \\
800      &  0.0620    &      25,366,688  &     0.0000  &    36,752   \\
2700      &  0.5770    &    290,570,648  &     0.0000  &    116,848   \\
6400      &  xx    &     xx   &     0.0160  &  322,192     \\
12500      &  xx    &    xx   &     0.0160  &  631,728    \\
21600      &  xx    &    xx   &     0.0310  &  995,728   \\
34300      &  xx    &    xx   &     0.0620  &  1,503,728  \\
51200      &  xx    &    xx   &     0.0940  &  2,310,800  \\ \hline
\end{tabular}
\caption{Left Recursive (lr) DCG: Space and Time Comparisons}\label{lrrec-table}
\end{center}
\end{figure}
Again note that interning makes it linear in time and space.

Figure \ref{darpa-avm-table} compares the space and time cost of
loading (and initializing) a large ontology when interning all ground
structures and when interning none.  
\begin{figure}
\begin{center}
\begin{tabular}{|c|c|c|c|c|c|} \hline
nonintern &  nonintern &   intern  &     intern &  intern &   intern \\
asserted space  &  cpu Time &    asserted space &   interned space & total space &  cpu Time \\
(bytes)	 &    (secs)  &    (bytes)    &     (bytes) &   (bytes) &   (secs)     \\ \hline
4,456,675,216 & 484.945 & 2,765,179,136 & 506,172,376 & 3,271,351,512 & 513.290 \\ \hline
\end{tabular}
\caption{Loading a large database of ontology facts}\label{darpa-avm-table}
\end{center}
\end{figure}
The application loads and
initializes a large ontology (and data) using XSB's CDF representation
(a package within the XSB System \cite{xsbmanual-3.3.x}).  The CDF
represents classes and objects and relationships between them.
Classes are represented by small terms, such as
cid(local\_class\_name,name\_space), and objects and properties
similarly.  Also measures (quantity and units) are represented by
small (and some not so small containing perhaps 15 symbols) ground
terms as well.  This ontology has over 1.7 million objects and 7.5
million attributes, represented by facts such as
hasAttr\_ext(Oid,Rid,Cid), where each id is an object id term,
relation id term, or class id term.  So there are many ground terms
asserted in this database, so interning ground subterms in these
asserted facts may save signification space.

When interning ground terms for all dynamic predicates,
there is 26.6\% decrease in space used traded for a 5.84\% increase in
load and initialization time.  Note that initialization includes more
than just the asserting of the facts; so the time overhead for just
the interning of asserted facts would be a higher percentage.  But
this does give an idea of the trade-offs of using interning in a large
and complex application.

Another perhaps nonintuitive example is the following simple program
that uses tabling and interning to provide asymptotic log access to
entries in a sorted list.

\begin{verbatim}
% find Ent in SortedList, which is Len long
find(Ent,Len,SortedList) :-  
    Len > 0,
    split_sorted(Len,SortedList,LoList,HiList),
    HiList = [Mid|_],
    (Mid == Ent
     -> true
     ;  LoLen is Len // 2,
        (Ent @< Mid
         -> find(Ent,LoLen,LoList)
         ;  HiLen is Len - LoLen, HiLen > 1,
            find(Ent,HiLen,HiList)
        ) ).

:- table split_sorted/4 as intern.
% Split a sorted list in half (knowing its length)
split_sorted(Len,List,LoList,HiList) :-
    Len1 is Len // 2, split_off(Len1,List,LoList,HiList).

split_off(Len,List,LoList,HiList) :-
    (Len =< 0
     -> LoList = [], HiList = List
     ;  List = [X|List1], LoList = [X|LoList1], Len1 is Len - 1,
        split_off(Len1,List1,LoList1,HiList)  ).

% Query to build a long list of 500,000 elements, 
% and look up 100 elements that are not there.
:- import intern_term/2 from machine.
?- mkevenlist(1000000,L0), cputime(T0),
   (intern_term(L0,L), between(1,I,100), I2 is 100*I+1,
    find(I2,500000,L), fail
    ;
    true  ),
   cputime(T1), Time is T1-T0, writeln(cputime(Time)), fail.
\end{verbatim}

This query builds a list of even numbers 500,000 elements long
starting from 0 and in increasing order.  It interns that ground list,
and then uses find/3 to use split\_sorted/4 to allow it to do a binary
search on the list to look up each of 100 odd numbers (of course
finding none of them.)  The basic work is done by split\_sorted/4,
which takes a sorted list and its length and produces two lists: the
first half of the list, and the second half, so the middle element of
the list is the first element in the second list, which split\_sorted/4
makes immediately accessible.  Since split\_sorted/4 is tabled as
intern, the lists that split\_sorted/4 generates are interned.  So no
explicit lists are stored in the trie, only pointers to interned
ground lists.  This query takes 0.2650 seconds, uses 111,464,680 bytes
of space for the interned terms and 466,608 bytes of table space to
store the calls to and returns from find/4.  Without the explicit call
to intern\_term/2, this query takes approximately the same space, but
over 7 seconds of cputime, since it has to intern the list of 500,000
elements for each of the 100 calls to find/3.  I didn't try running
this query without interning, for what are, I think, obvious reasons.

\section{Related Work}
Nguyen and Demoen \cite{nguyen-demoen-rep-sharing} describe the
general issue of sharing term representions in the implementation of
Prolog.  They motivate the advantages of representation sharing and
provide effective implementations.  They do not consider its potential
impact with respect to tabling.  They constrain their approach to
avoid any change in the standard representation of terms in their
implementations, while our approach does change the representation of
terms to the extent of interpreting tagged term references outside of
the heap to be interned terms.  Our unification algorithm does change,
minimally, to take advantage of the new representation.

Zhou and Have \cite{zhou-hash-consing} present an implementation of
hash-consing in B-Prolog with goals similar to those of this work: to
eliminate an unnecessary extra linear factor in both the time and
space complexity of tabling when naive copying of subgoals and answers
into and out of tables is done.  Their work is clearly prior to this,
but their algorithm is somewhat different, using hash tables instead
of tries to store tables and requiring and extra optimization of {\em
hash code memoization} to obtain the improved time complexity. It is
intimately connected with their implementation of tabling.  Our
algorithm can be effectively applied when asserting terms.  I was
motivated to write this paper, since I believe that this
implementation is simpler, clearer, and more general and orthogonal to
tabling, and deserves consideration as an alternative implementation.

\section{Discussion}
The approach to representation sharing described in this
paper is simple and designed primarily to improve the complexity of
tabling on certain kinds of programs.  The implementation in XSB is
usable, but further effort is necessary to make if fully robust.
Foremost, it must be extended to support the expansion of the size of
hash tables used to access the interned records.  This is
straightforward to do.  Secondly, I would like to support garbage
collection of the interned records.  XSB currently supports garbage
collection of the atom space.  It is not difficult to add to this
function the ability to garbage collect the interned space.

As described in Section \ref{int-bef-tab}, this implementaiton
currently don't allow the use of interned tables when interning might
compromise indexing.  However, there are cases, in particular when
using answer subsumption
\cite{answersubs:jelia},
in which loss of indexing would be acceptable.  I would like to
revisit and further explore this issue.  One option that suggests
itself, and might be good for other reasons, is to allow interning of
specific designated arguments, rather than interning all or none.

%

\bibliographystyle{plain}
\bibliography{bib}

%
%

\clearpage
\end{document}